\documentclass[epj,nopacs]{svjour}

\usepackage{amsmath,amssymb,graphicx,bm}
\usepackage{color}
\usepackage{multicol}        
\usepackage[bottom]{footmisc}

\newcommand{\beq}{\begin{equation}}
\newcommand{\eeq}{\end{equation}}
\newcommand{\bea}{\begin{eqnarray}}
\newcommand{\eea}{\end{eqnarray}}
\newcommand{\ba}{\begin{align}}
\newcommand{\ea}{\end{align}}
\newcommand{\bfig}{\begin{figure}}
\newcommand{\efig}{\end{figure}}
\newcommand{\cntilde}{\tilde{c}_n}
\newcommand{\D}{\displaystyle}
\newcommand{\rg}{\raggedright}
\newcommand{\gev}{\, \text{GeV}}
\newcommand{\mev}{\, \text{MeV}}
\newcommand{\thetaprime}{\theta^{\prime}}
\newcommand{\thetain}{\theta_{\rm in}}
\newcommand{\tpi}{t_{\pi}}
\newcommand{\tin}{t_{\rm in}}
\newcommand{\la}{\langle}
\newcommand{\ra}{\rangle}

\begin{document}


%
\title{Constraining the
low energy Pion electromagnetic form factor with space-like and phase of time-like data}
\author{B.\ Ananthanarayan \and S. Ramanan}

\institute{Centre for High Energy Physics,
Indian Institute of Science, Bangalore\ 560 012, India.\\ {\email{anant@cts.iisc.ernet.in} \\ \email{suna@cts.iisc.ernet.in}}}        

\date{\today}
%
\abstract{The Taylor coefficients $c$ and $d$ of the Pion EM form factor are constrained using analyticity, knowledge of the phase of the form factor in the time-like region, $4 m_\pi^2 \le t \le \tin$ and its value at one space-like point, using as input the ($g-2$) of the muon. This is achieved using the technique of Lagrange multipliers, that gives a transparent expression for the corresponding bounds. We present a detailed study of the sensitivity of the bounds to the choice of time-like phase and errors present in the space-like data, taken from recent experiments. We find that our results constrain $c$ stringently. We compare our results with those in literature and find agreement with the chiral perturbation theory results for $c$. We obtain ${\mbox d \sim {\cal{O}}(10)\gev^{-6}}$ when $c$ is set to the chiral perturbation theory values.
}
\titlerunning{Constraining $F_\pi(t)$ using space-like and phase of time-like data}
\authorrunning{B. Ananthanarayan \and S.Ramanan} 
\maketitle

\section{Introduction}
\label{introduction}

The Pion form factor continues to be of current interest~\cite{leutwyler,anant_ramanan,Sens-cilero,Raha}. In~\cite{anant_ramanan}, we developed a framework for obtaining constraints on the low-energy expansion coefficients of the Pion form factor using data from the space-like region ($t < 0$)~\cite{Bebek1,Brown,Bebek2,Amendolia,nucl-ex/0607007}, suitably extending an earlier work of Raina and Singh~\cite{RainaSingh}. It was shown using arguments of analyticity for the Pion form factor and a reliable estimate of the pionic contribution to the $(g-2)$ of the muon,  that it is possible to isolate allowed regions in the $c-d$ plane, where $c$ and $d$ are the Taylor coefficients in the low-energy expansion of the Pion EM form factor given as,
\beq
	F_\pi(t) = F_\pi(0) + \D\frac{1}{6} \la r^2_\pi \ra t + c t^2 + d t^3 + \cdots
\eeq
The form factor is analytic everywhere in the complex $t$ plane except for the cut along $(4 m_\pi^2, \infty)$.
Caprini~\cite{Caprini} has employed the phase of the time-like data along a part of the cut and the QCD polarization function $\Pi^{\prime}$ in order to constrain the coefficients ($c,d$). Each of the above independently use either pure space-like data or phase of time-like data, employing the method of Lagrange multipliers. In this work, we consider the problem of simultaneous inclusion of phase of time-like data and one space-like datum. The constraints are introduced through Lagrange multipliers: the constraints from the phase of time-like data is introduced through an Omn\`{e}s function, while the space-like constraint has a simple linear form as seen in~\cite{anant_ramanan}. Caprini also advances a method of implementing data from time-like phase as well as modulus that does not use the technique of Lagrange multipliers in~\cite{Caprini}.


As shown in~\cite{anant_ramanan,Caprini}, the coefficients constrained by the normalization of $F_\pi$ and the value of the pion charge radius $r_\pi$ satisfy the following inequality 
\beq
	\rg I = \sum_{n = 0}^{\infty} (c_n)^2 \le 1. 
	\label{ineq}
\eeq
The $c_n$'s are functions of the Taylor coefficients, $c$, $d$, and an ``outer function'' related to the observable of interest, {\it viz.}, ($g-2$) of the muon, $\Pi^{\prime}$ etc.
The inequality in Eq.~(\ref{ineq}), is modified by the additional constraints provided by the space-like data and the phase of time-like data along a part of the cut. We already saw in ref.~\cite{anant_ramanan}, the modification to Eq.~(\ref{ineq}) that comes through space-like constraints alone, where the constraints are introduced through Lagrange multipliers $\alpha_i$. Eliminating them results in a simple determinantal equation for the bounds. Here, we also take into account the phase of the form factor $\delta(t)$, for $t$ in the range $t_\pi$ to $t_{\rm in}$, where $t_{\rm in}= 0.8 \gev^2$. 
The Fermi-Watson final state theorem implies that in the elastic region
the phase of the form factor is given by the P-wave $\pi-\pi$ scattering
phase shift $\delta^1_1$.  For a detailed discussion of the theorem we refer to ref.~\cite{leutwyler}.
Analogous to~\cite{Caprini}, we vary the coefficients for $n \ge N+1$, assuming that the first $N$ coefficients are known through the normalization of $F_\pi$ and the value of $r_\pi$. We express the resulting constraints in a form where improvements, input by input, are manifest.  Setting $N=3$ and using an input for $I$ yields the ellipse in the $c-d$ plane. 

A good check for our results is obtained by setting $N = 2$ and comparing with the treatment of Bourrely and Caprini~\cite{CapriniBourrely}, who have considered a problem in the $\pi \,K$ sector and have included the $\pi \, K$ phase shift and constraints from the Callan-Treiman point, which is a time-like datum
in the analyticity domain.

This paper has been divided into the following sections. In Section~\ref{formalism}, we describe the Lagrange multiplier technique and obtain expressions for the constraints and bounds. In Section~\ref{data}, we discuss the model of the phase of time-like data we choose and the data sets we employ for the space-like part. We present our results in Section~\ref{results} and summarize in Section~\ref{conclusion}. 

\section{Formalism}
\label{formalism}  
\begin{figure*}[t]
	\begin{center}
	 \includegraphics*[angle = 0, width = 4.0in, clip = true]{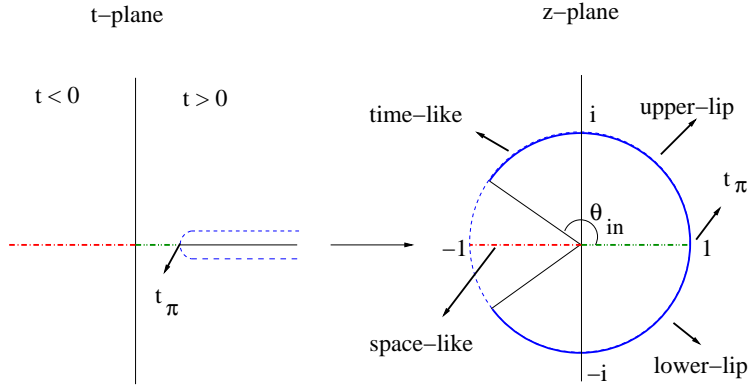}
	\caption{$z-t$ map for $z = e^{i\theta}$. The broken blue arc represents the time-like region, the dashed-dot red line the space-like region and the dashed-dot-dot-dashed green line the time-like region along positive real $t$ axis. The solid blue arc highlights the region in the $z$ plane which contributes to the phase data used here.}
	\label{phase_fig}
	\end{center}
\end{figure*}

The pion contribution to the muon anomaly is given by:
\beq
a_{\mu} (\pi^{+} \pi^{-}) = \D\frac{1}{\pi} \int_{t_{\pi}}^{\infty} dt\, 
\rho(t) |F_{\pi}(t)|^2
\label{amu.eqn1}
\eeq
where $t_\pi = 4 m_{\pi}^2$ is the branch point of the pion form factor and
\beq
\rho(t) = \D\frac{\alpha^2 m_{\mu}^2}{12 \pi} \frac{(t - t_\pi)^{3/2}}{t^{7/2}} K(t) \ge 0
\label{rho.eqn1}
\eeq
where,
\beq
K(t) = \int_0^1 du\, (1-u) u^2 (1- u + \D\frac{m_\mu^2 u^2}{t})^{-1}.
\label{rho.eqn2}
\eeq
We use the following map from the $t$-plane~\cite{anant_ramanan,RainaSingh}, that is cut from $t_\pi$ along the real $t$ axis, to the complex $z$-plane (region $|z|<1$),
\beq
\D \frac{z-1}{z+1} = i \sqrt{\frac{t - t_\pi}{t_\pi}}.
\label{ztmap.eqn1}
\eeq 
This map takes the point $t_\pi$ to $1$ and the point at infinity to $-1$, as seen in Fig.~(\ref{phase_fig}). Using this map and the definitions:
\beq
f(z) = F_{\pi}(t)
\label{defn.eqn1}
\eeq
\beq
p(z) = \rho(t)
\label{defn.eqn2}
\eeq
the pionic contribution to the muon anomaly can be written as
\beq
a_{\mu} (\pi^{+} \pi^{-}) = \D\frac{1}{2 \pi} \int_0^{2 \pi} d\theta \,w(\theta) |f(e^{i \theta})|^2
\label{amu.eqn2}
\eeq
where,
\beq
w(\theta) = 4 m_\pi^2 \sec^2 (\theta/2) \tan (\theta/2)\ p\,(e^{i\theta}) \ge 0. 
\label{amu.eqn3}	 
\eeq	
We now consider a function $h(z)$ defined as:
\beq
 h(z) = f(z) w_\pi(z)
\label{hz.eqn1} 
\eeq
where,
\beq 
w_\pi(z) = \exp\left[\D\frac{1}{4 \pi} \int_0^{2 \pi} d\theta\, \D\frac{e^{i\theta}+z}{e^{i\theta}-z} \ln w(\theta)\right].
\label{hz.eqn2}
\eeq
Then Eq.~(\ref{amu.eqn2}) can be written as:
\beq
a_{\mu}(\pi^{+} \pi^{-}) = \D\frac{1}{2 \pi} \int_0^{2 \pi} d\theta \,|h(e^{i \theta})|^2	
\label{amu.eqn4}
\eeq
Now $h(z)$ is analytic within the unit circle $|z| < 1$ and for real $z$, $h(z)$ is real. Therefore $h(z)$ can be expanded as follows:
\beq
h(z) = a_0 + a_1 z + a_2 z^2 + \cdots
\label{hz.eqn3}
\eeq
where $a_0, a_1 \cdots$ are real coefficients.
Therefore, in the analytic region, $|z| \le 1$, $a_{\mu}(\pi^{+} \pi^{-})$ can be written as:
\beq
a_{\mu} (\pi^{+} \pi^{-}) = a_0^2 + a_1^2 + \cdots
\label{amu.eqn5}
\eeq
a consequence of Parseval's theorem of Fourier analysis.
The expansion coefficients $a_n$ can be obtained from a Taylor expansion of 
the function $h(z)$ in terms of $f(z)$ and $w_{\pi}(z)$. The coefficients are 
given by
\beq
a_0 = h(0) = w_\pi(0),
\eeq
\beq
a_1 = h^{\prime}(0)= w_\pi^{\prime}(0) +\D\frac{2}{3} r_\pi^2 t_\pi w_\pi(0),
\eeq
\bea
a_2 & = &\D\frac{h^{\prime \prime}(0)}{2!}=\frac{1}{2}\left[
	w_\pi(0)\left(-\frac{8}{3} r_\pi^2 t_\pi + 32 \,c \, t_\pi^2\right) \right] \nonumber \\ 
	 &+& \frac{1}{2} \left[2 w_\pi^{\prime}(0)\left(\frac{2}{3} r_\pi^2 t_\pi\right) +w_\pi^{\prime \prime}(0) \right], 
\eea
and
\bea
a_3 &=& \D\frac{h^{\prime \prime \prime}(0)}{3!} =
\D\frac{1}{6} \left[ w_\pi(0)\left(12 r_\pi^2 t_\pi - 384\, c\, t_\pi^2 + 384 \,d \, t_\pi^3\right) \right] \nonumber \\
&+& \D\frac{1}{6}\left[3 w_\pi^{\prime}(0)\left(-\frac{8}{3} r_\pi^2 t_\pi + 32 \,c \, t_\pi^2\right)\right] \nonumber \\
&+&\D\frac{1}{6} \left[
2 w_\pi^{\prime \prime}(0) r_\pi^2 t_\pi + w_\pi^{\prime \prime \prime}(0) \right].  
\eea
Here, the expansion coefficients satisfy
\beq
\sum_{n = 0}^{\infty} (a_n)^2 = I
\label{cons.eqn2}
\eeq
Given $I$, Eq.~(\ref{cons.eqn2}) yields constraints on 
the expansion coefficients of
the form factor. 
Including up to the second (third) derivative 
for $F_{\pi}(t)$ results in constraints for $c$ ($c$ and $d$). Now if we divide Eq.~(\ref{cons.eqn2}) by $I$, then,
\beq
	\mu_0^2 = \sum_{n = 0}^{\infty} (c_n)^2 = 1
	\label{cons.eqn3}
\eeq
where,
\beq
	c_n = \D\frac{a_n}{\sqrt{I}}
	\label{cap_coeffn}
\eeq
Any $c_n$ which satisfies $\mu_0^2 \le 1$ is allowed and the equality gives the bound. We have already seen that this yields an ellipse in the $c-d$ plane~\cite{anant_ramanan,Caprini}.

The time-like phase is the argument of the form factor along a part of the cut. Let us assume that we know the phase in the region $\tpi \le t \le \tin$. We assume that, 
\beq
	{\rm Arg} \left[F_\pi(t+i\epsilon)\right] = {\delta}^1_1(t), \hspace*{0.2in} \tpi \le t \le \tin 
	\label{phase_cap_eqn1}
\eeq
in accordance with Watson's theorem, where $\delta^1_1(t)$ is the two-pion elastic scattering phase shift and $\tin$ $=$ $0.8 \,$ $\gev^2$ \,\cite{Caprini,Caprini_PRD}. 
The $z$ map, as defined in Eq.~(\ref{ztmap.eqn1}), maps $\tpi$ $\rightarrow$ $1$ $\Rightarrow$ $\theta$ $\rightarrow$ $0$ and $\tin$ $\rightarrow$ $z_{\rm in}$ $\Rightarrow$ $\theta$ $\rightarrow$ $\thetain$, where $z = e^{i\theta}$  The upper part of the cut is mapped on to the upper half circle and the lower part of the cut to the lower half circle. As a result,
\bea
	\rg {\rm Arg} \left[F_\pi(re^{i \theta})\right] &=& \delta^1_1(\theta), \hspace*{0.2in} 0 \le \theta \le \thetain \\
	\rg {\rm Arg} \left[F_\pi(re^{i \theta})\right] &=& -\delta^1_1(\theta), \hspace*{0.1in} (2\pi - \thetain) \le \theta \le 2 \pi 
	\label{phase_cap_eqn2}
\eea
as shown in Fig.~(\ref{phase_fig}).

The phase is introduced through the Omn\`{e}s function, denoted as ${\cal O}_\pi(z)$ in the $z$-plane, by,
\beq
	\rg {\cal O}_\pi(z) = \exp\left[\D\frac{i}{\pi} \int_0^{2 \pi} d\theta \frac{\bar{\delta}^1_1(\theta)}{1 - ze^{i\theta}} \right],
	\label{omnes_eqn1}
\eeq
where
\bea	
	\bar{\delta}^1_1 &=& \delta^1_1 \,\, {\rm if} \, \, 0 < \theta < \thetain \nonumber \\
			&=& -\delta^1_1 \, \, {\rm if} \, \, 2 \pi - \thetain < \theta < 2\pi. 
	\label{range_eqn2}
\eea
Since the phase of the form factor along the cut (i.e. $\tpi \le t \le \tin$) is compensated by the phase of the two pion scattering phase shifts, the following condition holds,
\beq
	\rg {\rm Im} \lim_{r \rightarrow 1}\left[\D\frac{1}{{\cal O}_\pi(r e^{i \theta})} f(r e^{i \theta}) \right] = 0 
	\label{time_like_cons1}
\eeq
which is the constraint equation coming from the time-like phase of the form factor.

In order to include information on phase of time-like data as well as space-like data, we consider the following Lagrangian
\bea
	\lefteqn{\rg {\cal L} = \D\frac{1}{2} \sum_{n = 0}^{\infty} c_n^2} \nonumber \\ 
	&& \mbox{}+ \D\frac{1}{\pi} \sum_{n = 0}^{\infty} c_n \lim_{\rm r \rightarrow 1} \int_{\Gamma} \lambda(\theta) |W(\theta)| {\rm Im} [[W(\theta)]^{-1} r^n e^{i n \theta}] d\theta \nonumber \\
	&& \mbox{} + \alpha (J - \sum_{n = 0}^{\infty} c_n z^n),
	\label{lag_eqn}
\eea
where,
\beq
	W(\theta) \equiv W(\zeta) = w_\pi(\zeta) {\cal O}_\pi(\zeta)
	\label{W_eqn}
\eeq
and 
\beq
	\zeta = \exp(i \theta).
\eeq
$J = h(z)/\sqrt{I}$ is the space-like datum mapped on to the $z$ plane as defined in Eq.~(\ref{ztmap.eqn1}) and $I$ is the bound from either $a_\mu$ or $\Pi^{\prime}$. We note that the space-like region is mapped on to the negative real axis as can be seen in Fig.~(\ref{phase_fig}). The value of $z$ and $h(z)$ used is given in Table~\ref{table_data}. As in~\cite{anant_ramanan,Caprini}, we use a value of $I = 75 \times 10^{-9}$ in the following.	 	
As mentioned in~\cite{anant_ramanan,RainaSingh}, the space-like constraints are linear and can be written as:
\beq
	\rg J - \sum_{n = 0}^{\infty} c_n z^n = 0
	\label{space_eqn1}
\eeq
and time-like constraints are given through the following equation~\cite{Caprini},
\beq
	\sum_{n = 0}^{\infty} c_n {\rm Im} \lim_{r \rightarrow 1}\left[[W(\theta)]^{-1} r^n e^{i n \theta} \right] = 0
	\label{time_eqn1}
\eeq
We minimize the Lagrangian in Eq.~(\ref{lag_eqn}) for $n \ge N+1$ where $N$ represents the terms already constrained using the normalization of $F_\pi$ and the pion charge radius $r_\pi$. In other words, we first constrain the coefficients though Eq.~(\ref{cons.eqn3}), truncating it at $N$ and then impose additional constraints through the time-like phase and space-like data. We have already seen that for $N = 3$, this yields an ellipse in the $c-d$ plane~\cite{anant_ramanan,Caprini}. Denoting the coefficients for $n \ge N+1$ as $\cntilde$ we have,
\bea
	\cntilde &=& -\D\frac{1}{\pi} \lim_{\rm r \rightarrow 1} \int_{\Gamma} \lambda(\theta) |W(\theta)| {\rm Im} [[W(\theta)]^{-1} r^n e^{i n \theta}] d\theta \nonumber \\
	&& \mbox{} + \alpha z^n \\ 
	&=& \D\frac{i}{\pi} \lim_{r \rightarrow 1} \int_{\Gamma} \lambda(\theta) \D\frac{|W(\theta)|}{W^*(\theta)} r^n e^{-i n \theta} d\theta + \alpha z^n
	 \label{const_eqn1}
\eea
Eqns.~(\ref{space_eqn1}) and~(\ref{time_eqn1}) can be written as follows:
\beq
	J - \sum_{n = 0}^N c_n z^n - \sum_{n = N+1}^{\infty} \cntilde z^n = 0
	\label{space_eqn2}
\eeq
\bea
	\sum_{n = 0}^N c_n {\rm Im} \left[\D\frac{e^{i n \theta}}{W(\theta)}\right] &+& \sum_{n = N+1}^{\infty} \cntilde {\rm Im} \lim_{r \rightarrow 1} \left[[W(\theta)]^{-1} r^n e^{i n \theta} \right] \nonumber \\
	&& \mbox{} = 0
	\label{time_eqn2}
\eea
In Eq.~(\ref{space_eqn2}), we substitute for $\cntilde$ from Eq.~(\ref{const_eqn1}) and solve for $\alpha$ so that,
\bea
	\alpha &=& \D\frac{(1 - z^2)}{(z^2)^{N+1}} \left[J - \sum_{n = 0}^{N} c_n z^n\right] + \D\frac{(1 - z^2)}{(z^2)^{N+1}} \nonumber \\
	 && \hspace*{-0.35in} \mbox{} \times \left[ \frac{i}{\pi} \int_{\Gamma} \lambda(\theta) \frac{|W(\theta)|}{W^*(\theta)} \frac{(z)^{N+1} e^{-i(N+1) \theta}}{1 - r e^{-i \theta}} d\theta \right].
	\label{alpha_eqn1}
\eea		
Next, we simplify the equation for time-like constraints (Eq.~(\ref{time_eqn2})), using $\cntilde$, given in Eq.~(\ref{const_eqn1}), so that we have,
\bea
	\lefteqn{\sum_{n = 0}^N c_n {\rm Im} \left[\D\frac{e^{i n \theta}}{W(\theta)} \right] + \alpha \sum_{n = N+1}^{\infty} z^n \lim_{r \rightarrow 1} {\rm Im} \left[\D\frac{r^n e^{i n \theta}}{W(\theta)} \right]} \nonumber \\
	&& \mbox{} - \D\frac{i}{\pi} \lim_{r \rightarrow 1} {\rm Im} \int_{\Gamma} \lambda(\thetaprime) \D\frac{|W(\thetaprime)|}{W^*(\thetaprime) W(\theta)} \sum_{N+1}^{\infty} r^n e^{in(\theta - \thetaprime)} d \thetaprime \nonumber \\ 
	&& = 0.
	\label{time_eqn3}
\eea
Using the following:
\beq
	W(\theta) = |W(\theta)| e^{i \Phi(\theta)}
	\label{wtheta_eqn1}
\eeq
where,
\beq
	\Phi(\theta) = \phi(\theta) + \bar{\delta}^{\,1}_1 (\theta),
	\label{phi_eqn1}
\eeq 
and $\phi(\theta)$ is the phase of the outer function $w_\pi(\zeta \equiv e^{i \theta})$, we substitute for $W(\theta)$ in Eq.~(\ref{time_eqn3}), that can be subsequently simplified to the following:
\bea
	\lefteqn{\sum_{n = 0}^N c_n \sin\left[n \theta - \Phi(\theta)\right] + \alpha \lim_{r \rightarrow 1} \sum_{n = N+1}^{\infty} {\rm Im} \left[(zr)^n e^{i (n\theta - \phi(\theta))} \right]} \nonumber \\
	&& \mbox{} - \D\frac{i}{\pi} \lim_{r \rightarrow 1} {\rm Im} \int_{\Gamma} \lambda(\thetaprime) \sum_{n = N+1}^{\infty} r^n e^{i \left[n (\theta - \thetaprime) - \Phi(\theta) + \Phi(\thetaprime) \right]} d \thetaprime \nonumber \\
	&& = 0.
	\label{time_eqn4}
\eea
Using Plemelj-Privalov relations~\cite{Caprini}, 
\beq
	\D\frac{1}{\pi} \lim_{r \rightarrow 1} \int_0^{2 \pi} \frac{F(\thetaprime)}{1 - r e^{i(\theta - \thetaprime)}} = F(\theta) + \frac{1}{\pi} {\cal P} \int_0^{2 \pi} \frac{F(\thetaprime)}{1 - e^{i(\theta - \thetaprime)}}
	\label{pp_eqn}
\eeq
and the expression for $\alpha$, given in Eq.~(\ref{alpha_eqn1}), we can simplify Eq.~(\ref{time_eqn4}) to get the following equation for $\lambda(\theta)$. 
\bea
	0 & = & \mbox{} -\lambda(\theta) + \sum_{n = 0}^N c_n \left[\sin(n \theta - \Phi(\theta))-\D\frac{1 - z^2}{z^{N+1}} \beta(\theta) z^n  \right] \nonumber \\
	&& \hspace*{-0.3in} \mbox{} + \D\frac{1}{\pi} \int_{\Gamma} d \thetaprime \lambda(\thetaprime) \frac{1}{2} \frac{\sin\left[(N+1/2) (\theta - \thetaprime) - \Phi(\theta) + \Phi(\thetaprime) \right]}{\sin\left[\frac{\theta - \thetaprime}{2}\right]} \nonumber \\
	&& \hspace*{-0.3in} \mbox{} + \D\frac{1}{\pi} \int_{\Gamma} d \thetaprime \lambda(\thetaprime) (1 - z^2)\beta(\theta) \beta(\thetaprime)  + J \D\frac{1 - z^2}{z^{N+1}} \beta(\theta)
	\label{lambda_eqn_main}
\eea
where 
\beq
	\beta(\theta) = \D\frac{\sin\left[(N+1) \theta - \Phi(\theta) \right] - z \sin\left[N \theta - \Phi(\theta) \right]}{1 + z^2 - 2 z \cos(\theta)} 
	\label{beta_eqn}
\eeq 
When no space-like constraints are used, Eq.~(\ref{lambda_eqn_main}), reduces to the $\lambda$ equation in~\cite{Caprini,micu}. We note that~\cite{micu} obtains the $\lambda$ equation keeping only the charge radius fixed.

We can further simplify the expression for $\alpha$, given in Eq.~(\ref{alpha_eqn1}), to get,
\beq
	\alpha = \D\frac{1 - z^2}{(z^2)^{N+1}} \left[J - \sum_{n = 0}^N c_n z^n + \frac{z^{N+1}}{\pi} \int_\Gamma d\thetaprime \lambda(\thetaprime) \beta(\thetaprime) \right]
	\label{alpha_eqn2}
\eeq
The simple condition $\mu_0^2 = \sum_0^{\infty} (c_n)^2 \le 1$~\cite{Caprini} now gets modified as follows 
\beq
	\mu_0^2 = \sum_{n = 0}^N (c_n)^2 + \sum_{n = N+1}^{\infty} (\cntilde)^2 \le 1.
	\label{mu_eqn1}
\eeq
We can once again substitute for $\cntilde$ and use Plemelj-Privalov relations~(\ref{pp_eqn}) to simplify the algebra so that,
\bea
	\mu_0^2 &=& \sum_{n = 0}^N (c_n)^2 + \D\frac{1}{\pi} \sum_{n = 0}^N c_n \int_{\Gamma} d \theta \lambda(\theta) \sin\left[n \theta - \Phi(\theta) \right] \nonumber \\
	&& \mbox{} + \alpha \left(J - \sum_{n = 0}^N c_n z^n \right) \le 1.
	\label{mu_eqn2}
\eea 

In our case, we will set $N= 3$ following~\cite{Caprini}. The region $\Gamma$ 
is defined as follows:
\beq
	\Gamma = \left[ \begin{array}{c}
	                \lbrace 0, \thetain \rbrace \\
			\lbrace 2 \pi - \thetain, 2 \pi \rbrace 
	               \end{array} \right].
	\label{range_eqn_gamma}
\eeq
From Eq.~(\ref{range_eqn2}), it is clear that information on the phase of the form factor along different parts of the cut can be used, i.e., $\bar{\delta}^1_1$ can be piece-wise continuous.

We solve for $\lambda(\theta)$ using eqns.~(\ref{lambda_eqn_main}), obtain the corresponding value of $\alpha$ from Eq.~(\ref{alpha_eqn2}). Using $c_n$'s,  already constrained by normalization of $F_\pi$ and pion charge radius, $\mu_0^2$ is evaluated (Eq.~(\ref{mu_eqn2})). Only those coefficients, $c_n$, which satisfy $\mu_0^2 < 1$ are retained.  

We have carried out a thorough check of our results by comparing with
similar results in the literature.  In particular,
our results for the case of $N = 2$ can be readily checked against results 
in~\cite{CapriniBourrely}, that were obtained for the case of $\pi \,K$ 
scattering, where the Callan-Treiman point was used. Although time-like, the Callan-Treiman point lies within the region of analyticity, i.e., $|z| \le 1$, and as a result, can be introduced through linear constraint equations.
\section{Data}
\label{data}
In this section, we present the data used to constrain $c$ and $d$. We use space-like data from recent experiments~\cite{Amendolia,nucl-ex/0607007}. These are summarized in~\cite{anant_ramanan}. We present the data here once again for the sake of completeness.

\begin{table*}
	\begin{center}
	\caption{Space-like data from Tadevosyan et al.~\cite{nucl-ex/0607007} and Amendolia et al.~\cite{Amendolia}, used in this work.}
	\label{table_data}
		\begin{tabular}{|l|l|l|l|l|}
		\hline\noalign{\smallskip}
		 & $t$($-Q^2$) [$\gev^2$] & $F_{\pi}(t)$ & $z(t)$ & $h(z) \times 10^{-5}$ \\
		\noalign{\smallskip}\hline\noalign{\smallskip}    	
		Tadevosyan & -0.600 & 0.433 $\pm$ 0.017 & -0.494 & 2.915\\ 
		Amendolia & -0.131 & 0.807 $\pm$ 0.015 & -0.242 & 4.454\\
		\noalign{\smallskip}\hline
		\end{tabular}
	\end{center}
\end{table*}

The time-like phase  we will use for part of this study is defined as
\beq
	\delta^1_1(t) = {\rm arc}\tan \left(\D\frac{m_\rho \Gamma_\rho(t)}{m_\rho^2 - t}\right) 
	\label{delta_eqn1}
\eeq
and
\beq
	\Gamma_\rho(t) = \D\frac{m_\rho t}{96 \pi f_\pi^2} \left(1 - \frac{4 m_\pi^2}{t} \right)^{3/2},
\eeq	
where $m_\rho = 770 \mev$ is the mass of the $\rho$ meson, $\Gamma_\rho = 150 \mev$ is the width of the $\rho$ resonance and $m_\pi = 139 \mev$ is the mass of the Pion. 
This parametrization first proposed in ref.~\cite{GP} and 
adopted in ref.~\cite{Caprini} has been
used here for ease of comparison of our results with those in
the literature. It may be noted
that at low energies, Eq.~(\ref{delta_eqn1}), agrees well with the one-loop chiral perturbation theory expression for the two-pion elastic scattering phase shifts and also with experiments for $t \ge 0.5 \gev^2$, as noted in~\cite{Caprini}.  
Therefore, we assume that the phase of the pion form factor coincides with Eq.~(\ref{delta_eqn1}) for $\tpi < t < \tin$, where $\tin = 0.8 \gev^2$. Our conformal map gives the following relation between $t$ and $\theta$ for $z \equiv \zeta = e^{i \theta}$
\beq
	t = \tpi + \tpi \tan^2(\theta/2)
	\label{t_theta.eqn1}
\eeq
from which we can calculate $\thetain$ for $\tin$.  

\begin{figure}
 \begin{center}
  \includegraphics[angle = 0, width = 3in, clip = true]{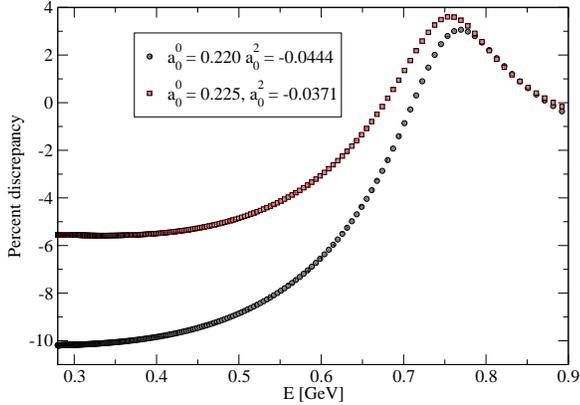}
  \caption{Percent discrepancy between the phases obtained from the analytical model used in~\cite{Caprini} ($\delta_\rho$) and those from the Roy equation using two different sets for the scattering lengths~\cite{ACGL,Colangelo:2001df}.}
  \label{Fig.phase_disc}
 \end{center}

\end{figure}

In order to test the sensitivity to the
parametrization of the phase, we adopt the phase shift obtained from Eq.(D1) of the
accurate Roy equation study of Ref.~\cite{ACGL}, for which the inputs are taken to be
$(a^0_0,a^2_0)$. Interestingly with the choice $(a^0_0,a^2_0)$ $= (0.225$, $-0.0371)$~\cite{ACGL} the phase
shifts do not differ from this analytical model used in~\cite{Caprini} by more than $5.5\%$ in the
entire range, where as with the more recent favored value of $(a^0_0,a^2_0)$ $= (0.220$, $-0.0444)$~\cite{Colangelo:2001df}, the
difference can be as much as $10\%$ as shown in Fig.~(\ref{Fig.phase_disc}). Despite this, the influence on our final results is not appreciable as we will show.

The Pion form factors can be directly determined by the K\"{u}hn-Santamaria and Gounaris-Sakurai parametrization ~\cite{barate}. We can, alternatively, use these two parametrization for the form factor, obtain the phase and check the sensitivity of the results obtained using Eq.~(\ref{delta_eqn1}). The central value of the parameters given in~\cite{barate} are used for the fits we consider in this work. As we will see, the general lack of sensitivity to the parametrization of the form factor, implies that no significant improvement will result from more recent or more precise inputs. 
 
In the next section we present our results for the above choice of parameters and data and also test the sensitivity of the bounds to variations in the data, choice of phase etc. 

\section{Results}
\label{results}

\begin{figure}[t]
 	\begin{center}
 	 \includegraphics*[angle = 0, width = 3in, clip = true]{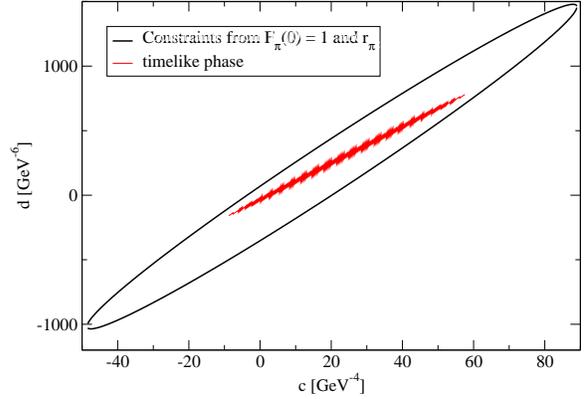}
	\caption{Results for constraining $c$ and $d$ using the normalization $F_\pi(0) = 1$ and the Pion charge radius, $r_\pi$ - large ellipse. We also see (small ellipse), the bounds obtained when the phase of the time-like phase alone is used in addition, using $a_\mu$ as input.}
	\label{nspace0_fig1}
	\end{center}
\end{figure}

\begin{figure}[t]	
	\begin{center}
 	\includegraphics*[angle = 0, width = 3in, clip = true]{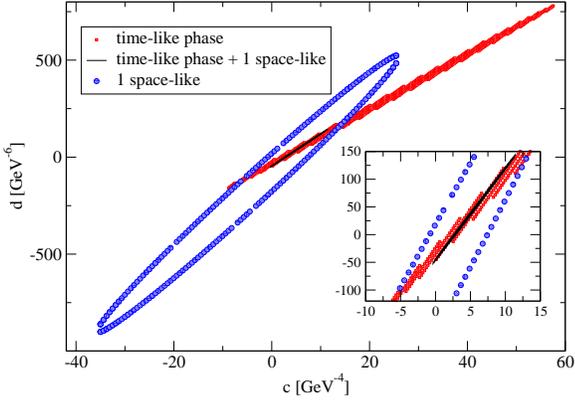}
	\caption{Results for constraining $c$ and $d$ using pure time-like (squares), pure space-like (circles) and combined time-like phase and one space-like datum (solid-line) from Tadevosyan et al (see Table~\ref{table_data}). The insert shows the area of intersection where the combined bounds lie.}
	\label{time-like_space_zoom.fig}
 	\end{center}
\end{figure}

In this section, we present our results for the bounds on the expansion coefficients and also check the sensitivity of the bounds to the errors in the input information.  
\begin{figure}[t]
 	\begin{center}
 	 \includegraphics*[angle = 0, width = 3in, clip = true]{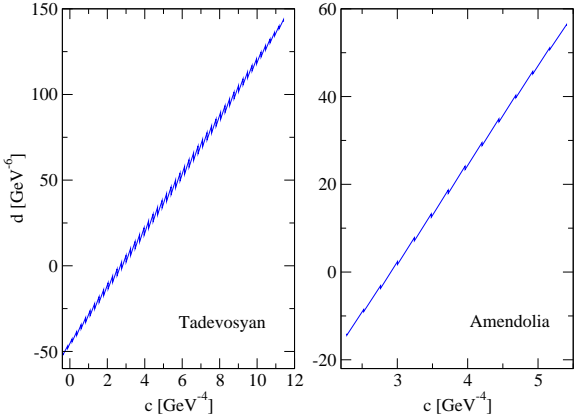}
	\caption{Results for constraining $c$ and $d$ using phase of time-like data along with one space-like data (value closest to $t = 0$), for the sets~\cite{Amendolia,nucl-ex/0607007}, as in table~(\ref{table_data}).}
	\label{time-like_space.fig}
 	\end{center}
\end{figure}

Fig.~(\ref{nspace0_fig1}) shows the bounds obtained using just the normalization of $F_\pi$ and the pion charge radius $r_\pi$ (large ellipse). The smaller ellipse, shows the improvements on the bounds when the phase of time-like data, as given in Eq.~(\ref{delta_eqn1}), introduced through the constraint Eq.~(\ref{time_like_cons1}), is used. The constraints obtained on the Taylor coefficients can be further improved by including the space-like data along with the phase of time-like data, discussed in section~(\ref{formalism}), as seen in Fig.~(\ref{time-like_space_zoom.fig}). The open ellipse is the allowed region in the $c$-$d$ plane when one space-like datum, from Tadevosyan et al (refer Table~\ref{table_data}), is used and the filled one represents that obtained when pure time-like phase is used. Now when both phase of time-like and a single space-like datum are combined, the allowed region is an ellipse in the region of intersection of the respective ellipses (refer inset in Fig.~(\ref{time-like_space_zoom.fig})). As is evident, the overlap region of the pure space-like and pure phase of time-like is significantly larger than the true region where they are taken simultaneously. 

We now focus on the bounds obtained when time-like phase and space-like datum are simultaneously used and address the sensitivity of these bounds to the variations in the input. Fig.~(\ref{time-like_space.fig}) shows the variation in the bounds obtained as the space-like datum moves away from $t=0$. The data point $h(z)$ from Amendolia, given in table~(\ref{table_data}), is closer to $t=0$ compared to the corresponding data point from Tadevosyan (table~(\ref{table_data}). The figure clearly shows that space-like data from low $t$ value constrain the coefficients better compared to data from higher $t$ value. We note that the same trend was observed for pure space-like constraints~\cite{anant_ramanan}.  
      
\begin{figure}[t]
 	\begin{center}
 	 \includegraphics*[angle = 0, width = 3in, clip = true]{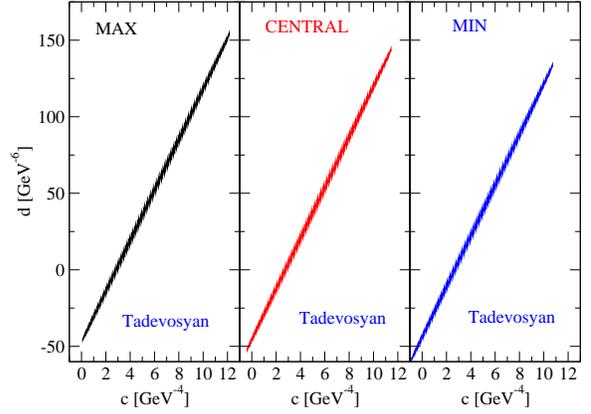}
	\caption{Variations in the bound as the space-like data from table~(\ref{table_data}) are varied within the error bounds. }
	\label{space_var.fig}
 	\end{center}
\end{figure}

\begin{figure}[t]
 	\begin{center}
 	 \includegraphics*[angle = 0, width = 2.75in, clip = true]{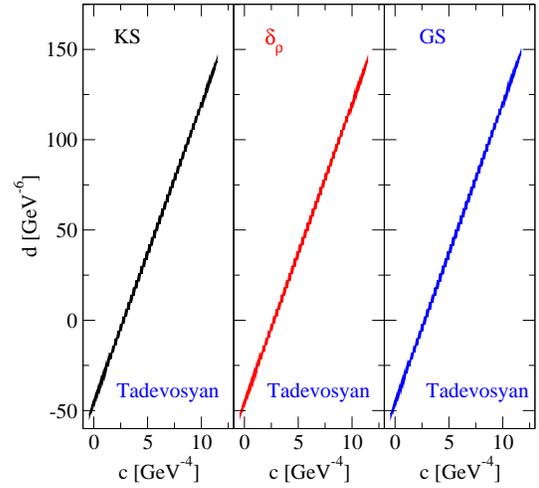}
	\caption{Variation in the bounds as the phase information is changed. $\delta_\rho$ is the parametrization for the $\delta^1_1(t)$ defined in Eq.~(\ref{delta_eqn1}).}
	\label{delta_rho_GS_KS.fig}
 	\end{center}
\end{figure}

\begin{figure}[t]
 	\begin{center}
 	 \includegraphics*[angle = 0, width = 3in, clip = true]{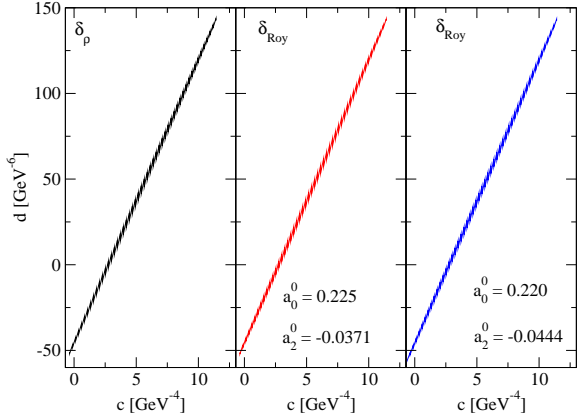}
	\caption{Comparing the constraints obtained when the fits given in Eq.~(\ref{delta_eqn1}) are compared to the more accurate ones from the Roy equation~\cite{ACGL,Colangelo:2001df}. Note that the dependence is negligible. The space-like datum comes from the set of Tadevosyan et al. (Table~\ref{table_data}).}
	\label{Fig.roy}
 	\end{center}
\end{figure}

\begin{figure}[t]
 	\begin{center}
 	 \includegraphics*[angle = 0, width = 3in, clip = true]{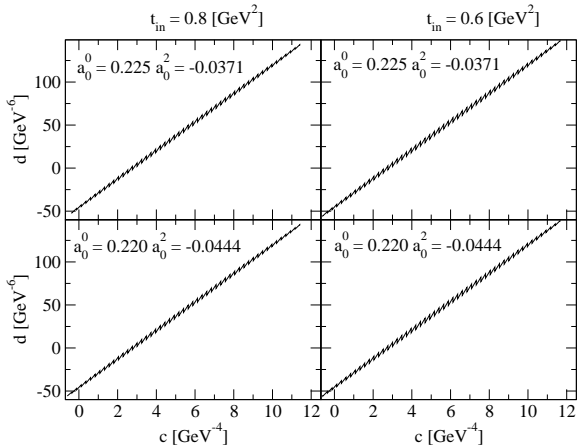}
	\caption{Varying the integration region for the phases obtained from the Roy equation~\cite{ACGL,Colangelo:2001df}. The left panel uses $\tin = 0.8 \gev^2$, while the right panel uses $\tin = 0.6 \gev^2$. Note the dependence on $\tin$, as well as on the two different parameter choices for the phase shifts from the Roy equation, is weak. }
	\label{Fig.roy_tin}
 	\end{center}
\end{figure}

\begin{figure}[ht]
 	\begin{center}
 	 \includegraphics*[angle = 0, width = 3in, clip = true]{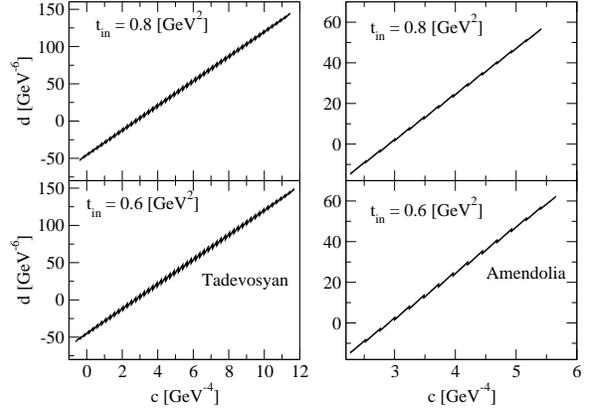}
	\caption{Variation in the bounds as the interval $\Gamma$ is varied. The phase is obtained from the $\delta_\rho$ model given in Eq.~\ref{delta_eqn1}. By changing $\tin$, more or less of the phase information is included.}
	\label{tin.fig}
 	\end{center}
\end{figure}

\begin{table*}
 	\begin{center}
	\caption{Bounds from phase of time-like data for different observables}
	\label{table_bounds_phase}
		\begin{tabular}{|l|l|l|}
		\hline\noalign{\smallskip}
		 & $\Pi^{\prime}$~\cite{Caprini} & $a_\mu$ \\
		\noalign{\smallskip}\hline
		$c \, [\gev^{-4}]$ & $[-14, 44]$ & $[-9, 58]$ \\
		$d \, [\gev^{-6}]$ & $[-236, 594]$ & $[-159, 780]$ \\
		\noalign{\smallskip}\hline
		\end{tabular}
 	\end{center}
\end{table*}


Fig.~(\ref{space_var.fig}) shows the variations in the bounds when the space-like datum in table~(\ref{table_data}) is varied away from its central value. Here, ``max'' refers to the central value $+$ maximum error and ``min'' is the central value $-$ maximum error. We see that the data is indeed sensitive to the errors in the space-like datum. However, varying the time-like phase between the definition in Eq.~(\ref{delta_eqn1}) and direct evaluation from K\"{u}hn-Santamaria (KS) or Gounaris-Sakurai (GS) fits for the form factor does not change the bounds on $c$ and $d$, as seen in Fig.~(\ref{delta_rho_GS_KS.fig}). This can be attributed to the fact that the phases produced by GS and KS fits and Eq.~(\ref{delta_eqn1}) agree with each other for the range of $t$ considered here. GS and KS parametrization involve two additional resonances compared to Eq.~(\ref{delta_eqn1}) that lie outside the range of $t$ used here. 

We have also carried out an analysis with accurate Roy equation fits to the
phase shifts. From Fig.~(\ref{Fig.roy}), we observe that the effect on the bounds of the coefficients $c$ and $d$ is rather weak as noted for other fits like KS or GS. The value of $\tin$, here fixed at $\tin = 0.8 \, \gev^2$, defines the interval $\Gamma$, as given in Eq.~(\ref{range_eqn_gamma}). Strictly speaking the solution (Eq.(D.1) of Ref~\cite{ACGL}) is valid up to 
$t_{\rm max}$ of $0.64 \gev^2$, but we assume that it is valid until $t_{\rm max} = 0.8 \gev^2$, as there is good agreement with the model phase $\delta_\rho$ (see Fig.~\ref{Fig.phase_disc}).  Furthermore, as shown in Fig.~(\ref{Fig.roy_tin}), lowering $\tin$ to $0.6 \gev^2$ does not lead to a perceptible 
change in the allowed region for $(c,d)$.  
Similarly, varying $\tin$ for the analytical model $\delta_\rho$ for the data sets from Amendolia and Tadevosyan has a very weak effect on the bounds, as in Fig.~(\ref{tin.fig}). 

\section{Discussion and Summary}
\label{conclusion}

In this paper, we study the improvements on the bounds of the low-energy Taylor expansion coefficients of the Pion EM form factor, when both the phase of the time-like data and one space-like datum is used. We use the method of Lagrange multipliers to include the constraints.
Bounds obtained using just the phase of time-like data with either $a_\mu$ or $\Pi^{\prime}$~\cite{Caprini} as input is shown in table~\ref{table_bounds_phase}. The results, when time-like phase and a single space-like datum are simultaneously used, are encouraging. In this work we
have carefully considered the effect of a variety of ways in which the
time-like phase is accounted for, by considering a $\rho$ dominant model,
phases from GS and KS parametrization of experimental data and accurate
Roy equation fits to scattering phase and have demonstrated that the
results are stable. Our best estimates are obtained when space-like data from the data set of Amendolia at al.~\cite{Amendolia}, as given in table~(\ref{table_data}), is used. In this case, the coefficient $c$ lies in the range: [$2.3 \gev^{-4}$, $5.4 \gev^{-4}$] and $d$ in [$-14 \gev^{-6}$, $56 \gev^{-6}$]. Furthermore, if the modulus of the form factor is included along with the phase~\cite{Caprini} and $\Pi^{\prime}$ as input, the $c$ and $d$ have the following ranges: [$0.5$ $\gev^{-4}$, $7.5$ $\gev^{-4}$] and [$-1$ $\gev^{-6}$, $22$ $\gev^{-6}$]. It is worth noting here, that the range of $c$ isolated in the present study, is the most stringent and is also in agreement with chiral perturbation theory, where $c$ has been determined to 2-loop accuracy: $4.49\, \gev^{-4}$~\cite{hep-ph/0203049}(see also ref.~\cite{gassermeissner}). 
As an interesting exercise, fixing the value of $c$ to that obtained from chiral perturbation theory, we read-off the range for $d$. Both time-like phase and time-like phase plus one space-like constraint gives $d$ in the range $\sim 20 - 30 \gev^{-6}$, when $a_\mu$ is used as input. Similar range is obtained for the bounds on $d$ when $\Pi^{\prime}$ and phase of time-like data are used as inputs. When the modulus of the form factor in the time-like region is also taken into account, along with the time-like phase information, the range of $d$ is roughly around $\sim 10 - 15$ $\gev^{-6}$. The slight mismatch between these ranges could warrant a separate study taking into account the uncertainties in parametrization of the phase and modulus and corrections to $\Pi^{\prime}$ and is clearly beyond the scope of this work. However it is remarkable that such a different variety of inputs and theory leads to a rather coherent picture for the values of $c$ and $d$. Finally it may be noted that the value of the bound on $d$ is an order of magnitude greater in this case, when pure space-like data is considered~\cite{anant_ramanan}  

The GS fit is well known for its good analytic properties. Therefore, the value of $c$ and $d$ can be obtained by expanding out the parametrization for the form factor. This gives an estimate for $c,d$ as $3.37 \gev^{-4}$ and $10.2 \gev^{-6}$ respectively and is well accommodated by our best constraints obtained for the Amendolia data set for the space-like part.
The value of $c$ so determined is consistent with another determination
available in the literature ref.~\cite{CFU} of $3.2\pm 0.5 \pm 0.9 \gev^{-4}$. The more accurate fits for the phase shifts from the Roy equation~\cite{ACGL,Colangelo:2001df} yield bounds which are very close to the ones obtained from the analytical model $\delta_\rho$. Fixing the upper limit of the integration region $\Gamma$ to be $\tin = 0.8 \gev^2$ and the space-like datum to that from Tadevosyan et al. (table~\ref{table_data}), the bounds on $c$ from the model $\delta_\rho$ for the phase shift is: [$-0.36$, $11.4$], which is identical to that obtained from the Roy equation fits using parameters from~\cite{ACGL,Colangelo:2001df}. The bounds obtained for $d$ varies slightly for different phase shift parameterizations, i.e., [$-51.12$, $142.51$] for the model phase shifts and [$-55.73$, $142.44$] for the Roy equation fits using parameters from~\cite{Colangelo:2001df}; while the range obtained for $d$ using the parameters from~\cite{ACGL} is identical to that obtained from the model phase shifts. 
From this we can conclude that the Roy solutions and the model $\delta_\rho$ essentially agree in the region of integration, especially in the resonance region.

At this point, it becomes important to include the modulus and phase of time-like data, along with space-like data, so that we include all the information available and constrain the pion form factor using arguments of analyticity. Hence, it would be interesting to explore the possibility of working along the lines of~\cite{Caprini}, where the modulus and phase of time-like data are used to obtain bounds, and combine it with the technique already explored in~\cite{anant_ramanan}. 

We note that the bounds are sensitive to the errors in the data used. Therefore, as an intermediate step, it would be worth-while to obtain a theoretical fit to the data ($\chi^2$ fit) and use the fit as input. This would eliminate the influence of experimental errors. 
In order to completely understand the sensitivity issues, it is important, as already noted in~\cite{anant_ramanan,RS_NPB}, to do a combined error analysis of time-like and space-like data, based on the work of Raina and Singh~\cite{RS_NPB}, that also gives the possibility of including the modulus via the technique of Lagrange multipliers. This theory needs to be developed.

\begin{acknowledgement}
 BA thanks DST, GOI for support. We thank I.~Caprini for discussions and comments. We thank H~Leutwyler for correspondence and Gauhar Abbas for assistance with some computational work. SR thanks G.~Colangelo and M.~Passera for useful discussions and ITP Bern and University of Padova for hospitality, where a part of this work was done. 
\end{acknowledgement}


\end{document}